\documentstyle[psfig,prl,aps]{revtex}
\begin{document}
\draft
\title{Quantum Nonlocality in Two-Photon Experiments at Berkeley}
\author{Raymond Y. Chiao\dag, Paul G. Kwiat\ddag~and Aephraim M. Steinberg\S}
\address{\dag\ Department of Physics,
University of California, Berkeley, CA 94720-7300, U.S.A.}
\address{\ddag\ Institut f\"ur Experimentalphysik, Universit\"at Innsbruck,
Technikerstrasse 25, A-6020 Innsbruck, Austria}
\address{\S\ National Institute of Standards and Technology,
Phys A167, Gaithersburg, MD 20899, U.S.A.}
\date{Preprint quant-ph/9501016; This version was produced on December 21,
1994}
\maketitle

\newcommand{\be}{\begin{equation}} \newcommand{\ee}{\end{equation}}
\newcommand{\mpm}{\mu_{\scriptscriptstyle +-}}

\begin{abstract}
We review some of our experiments performed over the past few years on
two-photon interference.  These include a test of Bell's inequalities, a study
of the complementarity principle, an application of EPR correlations for
dispersion-free time-measurements, and an experiment to demonstrate the
superluminal nature of the tunneling process.  The nonlocal character of
the quantum world is brought out clearly by these experiments.  As we explain,
however, quantum nonlocality is not inconsistent with Einstein causality.
\end{abstract}

\section{Introduction}

We shall begin with a brief review of the Einstein-Podolsky-Rosen (EPR)
``paradox'' \cite{Einstein=1935}, and then review some of our experiments at
Berkeley:  the Franson experiment \cite{Franson=1989,Kwiat=1993FR}, the
``quantum eraser'' \cite{Kwiat=1992}, the ``dispersion-cancellation'' effect
\cite{Steinberg=1992PRL}, and tunneling-time measurements
\cite{Steinberg=1993PRL,Steinberg=1994SUB}.  Let us begin by stating that we
consider the EPR phenomenon to be an ``effect,'' not a ``paradox'': EPR's
experimental predictions are internally consistent, and a contradiction is only
reached if one assumes both EPR's notion of locality and the completeness of
quantum mechanics (QM).  The three central elements that constitute the EPR
argument are 1) a belief in some of the quantum-mechanical predictions
concerning two separated particles, 2) a very reasonable definition of an
``element of reality'' [namely, that ``if, without in any way disturbing a
system, we can predict with certainty (i.e., with probability equal to unity)
the value of a physical quantity, then there exists an element of physical
reality corresponding to this physical quantity''], and 3) a belief that nature
is local, i.e., that no expectation values at a spacetime point $x_2$ can
depend on an event at a spacelike-separated point $x_1$ (this definition of
locality is now seen to be more stringent than Einsteinian causality, and is
inconsistent with QM).  In the original EPR scheme \cite{Einstein=1935}, the
system under consideration is a pair of particles described by the wavefunction
$\delta({x_{1}-x_{2}-{\rm a}})$.  This is a simultaneous eigenstate of the
operator $(X_{1}-X_{2})$ [with eigenvalue a], and the operator $(P_{1}+P_{2})$
[with eigenvalue 0], which is possible since the commutator
$[X_{1}-X_{2}, P_{1}+P_{2}]$ vanishes.  In other words, the {\it sum} of the
particles' momenta is well-defined, as is the {\it difference} of their
positions.  Therefore, if we measure the momentum of one particle, we can
predict with certainty the momentum of the other (hypothesis 1), even though it
may be sufficiently far away that no signal could be transferred between them
(hypothesis 3).  Momentum must therefore be an element of reality
(hypothesis 2).  And if we measure the position of one, then we can predict
with certainty the position of the other (possibly distant) particle.  Hence,
position is also an element of reality.  But quantum mechanics does not allow a
precise specification of both the position and the momentum of a particle.
This was not a paradox to EPR; rather, they concluded that quantum mechanics
must be incomplete--a complete theory would not contradict the predictions of
QM, but would incorporate all elements of reality, just as atomic theory
incorporates the positions and momenta of individual atoms, variables not
described by thermodynamics.

\begin{figure}
\centerline{\psfig{width=7.5cm,figure=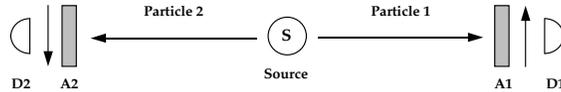}}
\caption{Bohm's version of the EPR Gedankenexperiment}
\label{eprexp}
\end{figure}
\begin{figure}
\centerline{\psfig{width=7.5cm,figure=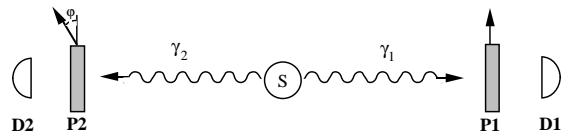}}
\caption{Optical version of EPR experiment}
\label{opvern}
\end{figure}
In Bohm's version of the EPR effect \cite{Bohm=1983}, (see
Fig.~\protect\ref{eprexp})
a spin-0 particle decays into two spin-1/2 particles in the singlet state
\begin{equation}
|\rm{Singlet}\rangle=\frac{1}{\sqrt{2}}\{|\uparrow_{1}\rangle|\downarrow_{2}
\rangle-|\uparrow_{2}\rangle|\downarrow_{1}\rangle\}\;.
\label{Bohm}
\end{equation}
The filters A1 and A2 in Fig.\,\protect\ref{eprexp} are Stern-Gerlach
apparatuses.
Optical versions of this experiment performed by Freedman and Clauser, and by
Aspect {\it et~al}., used photons in place of the spin-1/2 particles, and
linear polarizers (P1, P2) in place of Stern-Gerlach apparatuses
\cite{Freedman=1972,Clauser=1978,Aspect=1982ADR}; see
Fig.~\protect\ref{opvern}.
The coincidence count rate as a function of the relative angle between
polarizers P1 and P2 is a measure of the correlated behavior of the two
separated particles.  Bell derived an inequality \cite{Bell=1964} starting
from EPR's very general and seemingly reasonable notions of locality and
realism; this inequality is violated by the high-visibility sinusoidal fringes
predicted by quantum mechanics.  Most importantly, {\it experiments} yield
results in agreement with quantum mechanics, and in violation of this
inequality (modulo some reasonable auxiliary assumptions about uncounted
particles).  Therefore, these experiments rule out a broad class of local
realistic theories.

These early experiments relied on the correlations of the polarization (i.e.,
{\it internal} degrees of freedom) of the particles.  Since the predictions of
quantum mechanics are so strange, it is important to investigate them for
{\it external} degrees of freedom as well, such as the momentum and position of
the particles considered in the original EPR paper.  Rarity and Tapster have
already done so for momentum and position \cite{Rarity=1990BI}, and recently we
have performed an experiment \cite{Kwiat=1993FR} (first proposed by Franson
\cite{Franson=1989}) relying on the external variables of energy and time.

\section{Entangled states}

Erwin Schr\"{o}dinger \cite{Schrodinger=1983}, in response to the EPR paper,
pointed out that at the heart of these nonlocal effects is what he called
``entangled states'' in quantum mechanics, i.e., {\it nonfactorizable}
superpositions of product states.  For if a two-particle wavefunction were
factorizable,
\begin{equation}
\psi(x_{1},x_{2})=\chi(x_{1})\chi(x_{2})
\end{equation}
then the probability of joint detection would also factorize,
\begin{equation}
|\psi(x_{1},x_{2})|^{2}=|\chi(x_{1})|^{2}|\chi(x_{2})|^{2}
\end{equation}
so that the outcomes of two spatially separated measurements would be
{\it independent} of one another.  In cases where the two-particle state cannot
be factorized as above, this means that the quantum-mechanical prediction
implies nonlocal correlations in the behavior of remote particles.   The Bohm
singlet state (\protect\ref{Bohm}) is such an entangled state.  It predicts
correlations between spin measurements made on the two particles. But these
correlations persist even if the particles and their analyzers are separated by
space-like intervals, implying the existence of non-local influences.  Though
each particle considered individually is unpolarized, the two particles will
{\it always} have opposite spin projections when measured along the same
quantization axis.  Einstein {\it et~al}. would conclude that each spin
component is an ``element of reality'' in that it would be possible to predict
its value with 100\% certainty without disturbing the particle, simply by
measuring the corresponding spin component of the particle's twin (a
measurement which according to EPR's locality hypothesis cannot disturb the
particle in question).  As discussed above, this reasoning led EPR to conclude
that quantum mechanics was incomplete; if one instead considers QM to be a
complete theory, one must then admit the existence of nonlocal effects.  As we
shall see below, experiment supports this latter interpretation.


\begin{figure}
\centerline{\psfig{width=7.5cm,figure=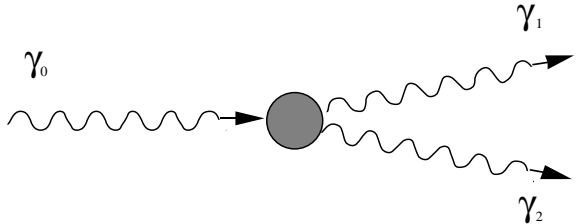}}
\caption{Two-photon decay from one photon}
\label{2phdcay}
\end{figure}
In our experiments, the entangled state we start with is the
{\it energy-entangled} state of two photons produced in a two-photon decay
process known as spontaneous parametric fluorescence.  The Feynman diagram for
this process is shown in Fig.~\protect\ref{2phdcay}, and the state of the
two-photon
system after decay from a parent photon with a sharply defined energy $E_0$ is
given by
\begin{equation}
|2\,\rm{photons}\rangle=\int_{0}^{E_0}\int_{0}^{E_0}dE_{1}dE_{2}
\delta(E_{0}-E_{1}-E_{2})
A(E_{1},E_{2})|E_{1}\rangle|E_{2}\rangle\,.\hspace{1.9cm}
\label{tang}
\end{equation}
Instead of a sum, as in the singlet state, we now have an integral, since
energy is a continuous variable.  The meaning of this energy-entangled state is
that after the measurement of one photon's energy gives the sharp value
$E_{1}$, there is an instantaneous ``collapse'' to the state
\begin{equation}
|E_{1}\rangle|E_{0}-E_{1}\rangle.
\end{equation}
This effect has been seen in an earlier experiment \cite{Chiao=1991MD}, in
which coincidences were recorded between photon $\gamma_1$, which passed
through an interference filter (to measure its frequency, and hence its energy,
with high resolution) and photon $\gamma_2$, which passed through a Michelson
interferometer (to measure its width); when photon $\gamma_1$ was detected
after the narrow-band filter, photon $\gamma_2$ collapsed into an equally
narrow-band energy state, whose coherence length was then far greater than that
of the uncollapsed state.

\section{The two-photon light source}

\begin{figure}
\centerline{\psfig{width=7.5cm,figure=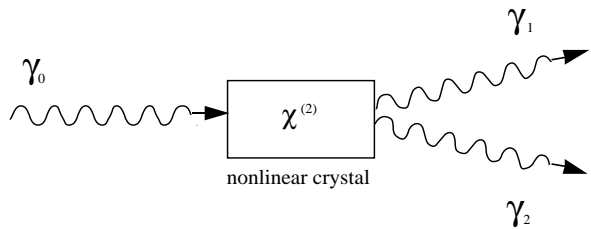}}
\caption{Parametric down-conversion}
\label{paradown}
\end{figure}
\begin{figure}
\centerline{\psfig{width=8cm,figure=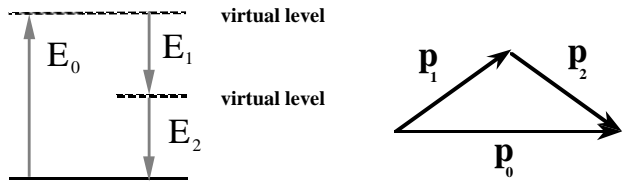}}
\caption{Energy level diagram; momentum conservation triangle}
\label{e&pparadown}
\end{figure}
\begin{figure}
\centerline{\psfig{width=8.2cm,figure=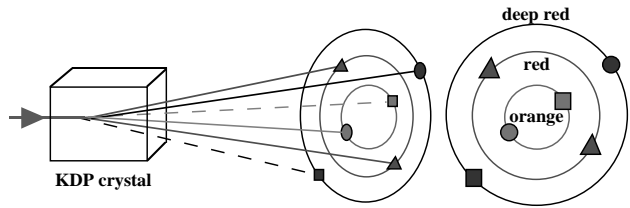}}
\caption{Parametric down-conversion process.  Matching shapes
represent conjugate photons, while each ring represents a different color.}
\label{parashape}
\end{figure}
The two-photon decay, or parametric down-conversion, occurs inside a crystal
with a $\chi^{(2)}$ nonlinearity (we used a potassium dihydrogen phosphate
(KDP) crystal).  This process is the reverse of second harmonic generation, in
which two red photons combine to form an ultraviolet photon at twice the
frequency.  In our crystal, a single photon $\gamma_0$ produced by a cw
ultraviolet laser (a single-mode argon ion laser at 351 nm) down-converts to
two red photons $\gamma_1$ and $\gamma_2$, each near 702 nm.  Energy and
momentum are conserved here:
\begin{eqnarray}
E_{0}&=&E_{1}+E_{2}\nonumber\\
\bf{p}_{0}&=&\bf{p}_{1}+\bf{p}_{2}\;.
\end{eqnarray}
The parent photon $\gamma_0$ is called the ``pump'' photon, daughter photon
$\gamma_1$ the ``signal'' photon, and daughter photon $\gamma_2$ the ``idler''
photon, for historical reasons.  A rainbow of colored cones (see
Fig.~\protect\ref{parashape}) is produced around an axis defined by the UV
laser beam,
where any two correlated photons will lie on opposite sides of the cone, e.g.,
the inner ``square'' orange photon is conjugate to the outer ``square''
deep-red photon, etc.  The two conjugate photons are always produced
essentially simultaneously in the two-photon decay:  They have been observed to
be born within tens of femtoseconds of each other.  Though due to the
continuous-wave nature of the pump laser, no expectation values of this source
are time-dependent, the photon correlations mean that once one particle is
detected, the other collapses into a very short wavepacket.  The wavepacket is
narrow because there are many ways to partition the energy of the parent
photon, so that each daughter photon has a broad spectrum.  However, due to the
entanglement [described by the state (\protect\ref{tang})], the sum of the two
photons'
energies is extremely well-determined.  Thus, the {\it sum} of their energies
and the {\it difference} of their arrival times can be simultaneously known to
high precision, though the absolute time of emission is unknown.  This is to be
contrasted with the case of the single particle, whose time and energy may not
be known to arbitrary accuracy due to the uncertainty principle; hence the
correlations are of the exact sort Einstein {\it et~al} used to argue that
quantum mechanics must be incomplete.

\section{The Franson Experiment}

\begin{figure}
\centerline{\psfig{width=16cm,figure=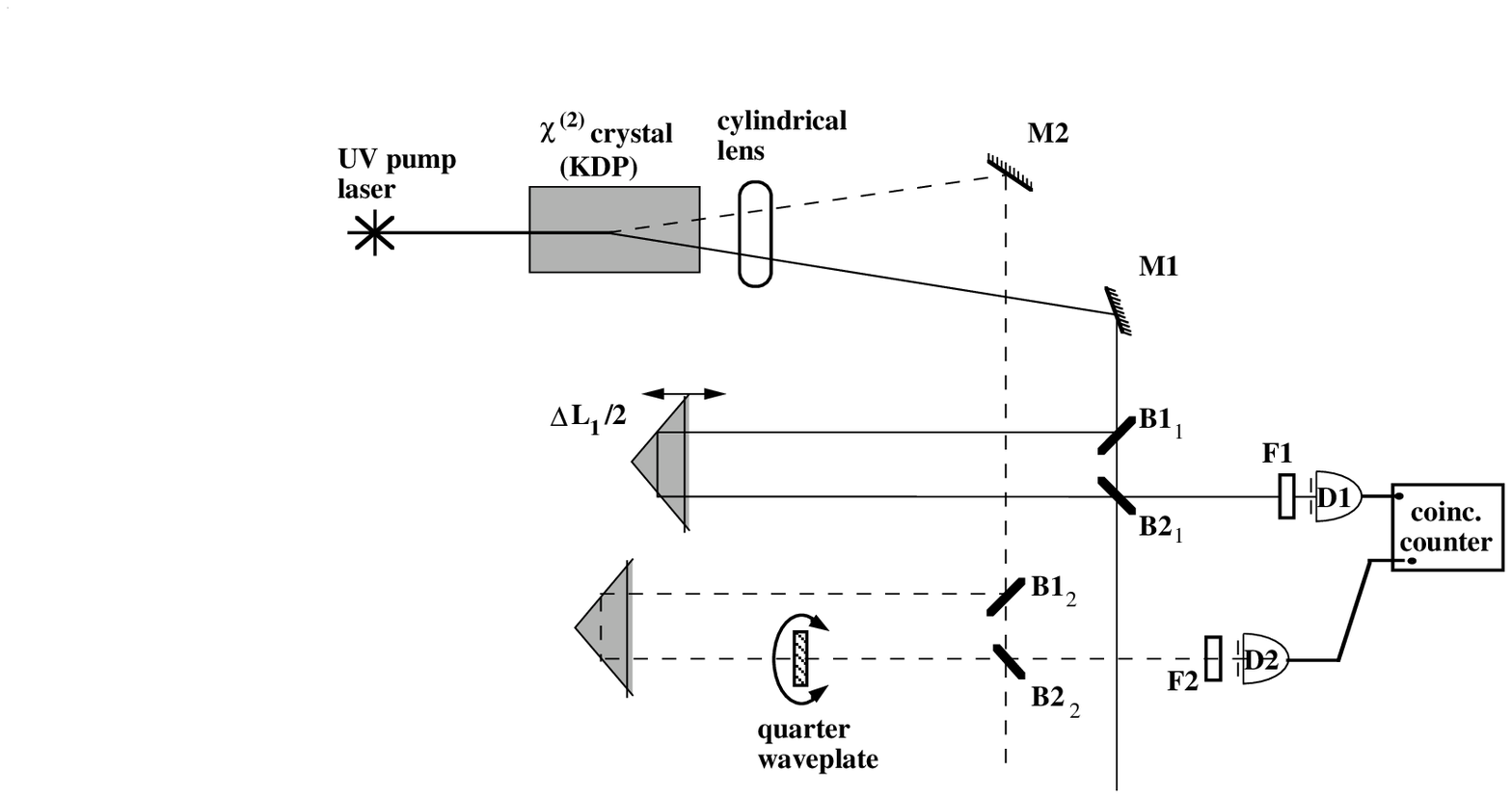}}
\caption{Apparatus used at Berkeley to perform the Franson experiment}
\label{berkfran}
\end{figure}
The Franson experiment \cite{Franson=1989} allows an examination of Bell's
inequality for energy and time, a violation of which implies that photons do
not necessarily possess a well-defined energy nor a well-defined time of
emission before measurement.  Our experiment \cite{Kwiat=1993FR} is sketched in
Fig.~\protect\ref{berkfran}. (Similar experiments were performed by Brendel
{\it et~al}
\cite{Brendel=1992}, and by Rarity and Tapster \cite{Rarity=1994}.)  The
correlated photons are directed to two separate but identical Mach-Zehnder-like
interferometers, in which they are allowed to take either a short path or a
long path.  These interferometers can have their path-length differences
adjusted by means of trombone prisms in their long paths and a waveplate in
that of interferometer 2.

There is no first-order interference inside a single interferometer, because
the coherence lengths of the photons ($\sim50~\mu$m) are much smaller than the
63-cm path-length differences in the interferometers.  However, there is
second-order (i.e., two-photon) interference observable in coincidence
detection between detectors D1 and D2.  We shall use Feynman's rules for
interference to calculate the probability of coincidence detection.  The
indistinguishable processes here are (1) the ``short-short" and (2) the
``long-long" processes (where both photons take their interferometers' short or
long paths, respectively).  The distinguishable processes are (3) the
``short-long" and (4) the ``long-short'' processes, since the delays between
the ``clicks" of D1 and D2 differentiate these events from one another, as well
as from processes (1) and (2).  In principle and also in practice, we are able
to reject these distinguishable ``clicks" by using a sufficiently narrow
coincidence timing window in our electronics.  We are thus left with only the
two indistinguishable processes (1) and (2), for which we must first add the
probability {\it amplitudes}, and then take the absolute square.  Hence the
probability of a given coincidence detection is given by the expression
\begin{eqnarray}
P_c&\propto&|1{\cdot}1-e^{i\phi_1}{\cdot}e^{i\phi_2}|^2\nonumber\\
&{\propto}&[1-\cos(\phi_{1}+\phi_{2})]\;,
\label{propto}
\end{eqnarray}
where the first term inside the absolute value corresponds to the
``short-short'' process, and the second term to the ``long-long'' process.
(The beam splitters are assumed to be 50/50 throughout, and a $\pi/2$ phase
shift for reflection is assumed.)  Here the phases $\phi_1$ and $\phi_2$
represent the phase differences between the short and long arms of
interferometers 1 and 2, respectively.

Note that equation~(\protect\ref{propto}) implies a fringe visibility of 100\%,
i.e.,
perfect zeros at the minima in coincidence detection.  The meaning of these
minima is that the two spatially separated photons behave in an
{\it anticorrelated} fashion: if one is transmitted at the final beam splitter,
then the other is reflected.  The meaning of the maxima of
equation~(\protect\ref{propto}) is that the two photons behave in a {\it
correlated}
fashion: either both are transmitted at the final beam splitter, or both are
reflected.  The behavior of the twins depends on the settings of the separate
phase shifters; the phases in our experiment were varied continuously, using a
piezo-electrically driven trombone arm in one interferometer ($\phi_1$), as
well
as discretely in increments of $90^{\circ}$, using a quarter waveplate in the
other interferometer ($\phi_2$).  Note that in principle the phases could even
be set {\it after} the photons had entered the interferometers.

\begin{figure}
\centerline{\psfig{width=8.5cm,figure=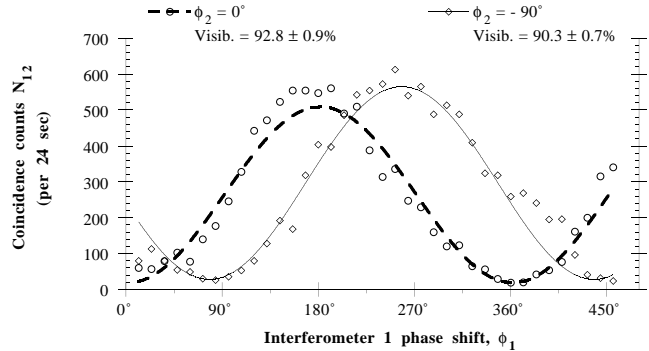}}
\caption{Interference fringes of our Franson experiment, for different phase
settings of the quarter waveplate: $\phi_2\,=\,-90^{\circ}\,(\diamond)$;
$\phi_2\,=\,0^{\circ}\,(\circ)$}
\label{fringe2}
\end{figure}
In  Fig.~\protect\ref{fringe2} we present our latest data.  The mere existence
of
interference fringes means that one cannot know, even in principle, the actual
emission times of the photons, for otherwise their arrival times at the
detectors would indicate whether the long or short paths had been taken, and
there would no longer be two interfering processes (see the following
section).  Bell's inequality for this experiment is violated by fringes which
vary sinusoidally with $\phi_1\,+\,\phi_2$ and have visibility greater than
$70.7\% (=1/\sqrt{2})$, once certain reasonable assumptions about uncounted
photons are made.  In our experiment, we observed fringes with the expected
dependence only on the sum of the phases, with visibilities exceeding the limit
by as many as 28 standard deviations.  Using a ``fair-sampling'' assumption and
the symmetry properties of the interferometer (in particular, that the
coincidence rate of the unused ports in Fig.~\protect\ref{berkfran} is equal to
that of
the used ports, an assumption supported by tests done with a third detector not
shown in the figure), we can directly obtain the
value of the Bell-parameter $S$ from the coincidence rates obtained at two
values each of $\phi_1$ and $\phi_2$.  $S$ is a measure of the strength of the
correlations between the two particles, evaluated for the four combinations
of the two values of $\phi_1$ and $\phi_2$, and according to the
Clauser-Horne-Shimony-Holt form of Bell's inequality \cite{Clauser=1969},
satisfy $|S|\,\leq\,2$ for any local realistic model.  For
appropriate choices of $\phi_1\,(45^{\circ}$ and $135^{\circ})$ and
$\phi_2\,(0^{\circ}$ and $-90^{\circ}$), we obtain  $S\,=\,-2.63\,\pm\,0.08$,
clearly displaying quantum nonlocality \cite{Kwiat=1993TH}.

\section{The ``Quantum eraser''}

\begin{figure}
\centerline{\psfig{width=7.2cm,figure=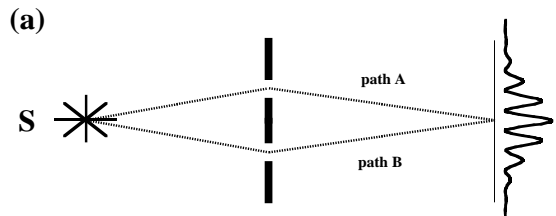}}
\caption{(a) Young's two-slit experiment, (b) with circular polarizers,
(c) with circular polarizers and linear polarizer; for latex reasons
temporarily beyond our control, b and c removed}
\label{2slit}
\end{figure}
Although the nonlocality of quantum mechanics is most apparent in tests of
Bell's inequalities, it also plays a central role in experiments exploring
complementarity.  One such, the ``quantum eraser,'' was discussed by Scully,
Englert and Walther \cite{Scully=1991} in connection with the micromaser.  In
Fig.~\protect\ref{2slit} we first present a simple precursor to this idea.
Consider
Young's two-slit experiment from a particle viewpoint.  The reason we see
interference at the screen in Fig.~\protect\ref{2slit} (a) is that one cannot
know,
{\it even in principle}, which path (A or B) the particle took on its way to
the screen.  This lack of which-path information is fundamental to the
observability of interference fringes.  However, suppose we placed two circular
polarizers of opposite senses, CP1 and CP2, in front of the two slits, as shown
in Fig.~\protect\ref{2slit} (b).  The photons which have passed through these
circular
polarizers are now {\it labeled} by their polarizations, so that by measuring
their helicities, one could tell which path the photons took to the screen.
Hence we shall call these polarizers ``labelers.''   Since we now have
which-path information, the interference pattern on the screen disappears; half
the time, the diffraction pattern of slit 1 appears, and half the time the
diffraction pattern of slit 2 appears, but since these two processes are made
mutually exclusive by the which-path labels, no interference occurs between
them.  (Note that the center-of-mass motion of the particles is in no way
disturbed in the transverse directions by the insertion of the circular
polarizers, so that this scheme is very different from Feynman's
\cite{Feynman=1965}, where due to the uncertainty principle, the scattering of
a particle near one of the slits uncontrollably disturbs the transverse
center-of-mass motion.  However, Stern {\it et al} \cite{Stern=1990} have
proved that the presence of a which-path detector in an interferometer always
leads to a random phase shift, thus washing out the interference fringes.)  Now
let us ``erase'' the which-path information by the insertion of a linear
polarizer LP in front of the screen; see Fig.~\protect\ref{2slit} (c).  The
linear
polarizer erases the handedness of the photons, which served as their labels.
(Note that the orientation of the axis of this linear polarizer is immaterial
for this erasure to occur; it only affects the phase of the resulting
interference pattern.)  Since which-path information is no longer available,
the interference pattern is revived.

\begin{figure}
\caption{Figure temporarily unavailable.
Hong-Ou-Mandel interferometer; resulting coincidence dip.}
\label{hongman}
\end{figure}
This particular version of the quantum eraser has a straightforward
classical-wave explanation when the light source is describable in terms of
coherent states. Thus it could be argued that there is nothing particularly
quantum about this quantum eraser.  Nevertheless, Jordan has proposed a similar
Mach-Zehnder version of this experiment \cite{Jordan=1993}, in which he has
argued on the basis of the correspondence principle that despite the existence
of a classical explanation, such first-order interference experiments can be
interpreted as true quantum erasers.  However, in order to avoid any possible
ambiguity concerning the quantum versus classical interpretation, we decided to
use the nonclassical two-photon light source described above, in conjunction
with the Hong-Ou-Mandel (HOM) two-photon interferometer \cite{Hong=1987}, to
demonstrate a quantum eraser with no classical analog.  Due to the use of
single particles, the meaning of which-path information is clear, whereas
in a classical-wave experiment, it is questionable whether there is any way to
ask which path a wave took.  In the interferometer shown in
Fig.~\protect\ref{hongman},
the two twin photons are brought back together by means of mirrors, so that
they impinge simultaneously on a 50/50 beam splitter, after which they continue
on to the two detectors D1 and D2.  The coincidence rate recorded by these
detectors is observed to go through a sharp dip as the path length difference
between the two photons is varied.  The width of this dip in our experiments is
typically $\stackrel{<}{_\sim}\!100$ femtoseconds, allowing very high
resolution in time-of-flight comparisons between the two photons, as we shall
discuss in detail in sections \protect\ref{1.6} and \protect\ref{1.7}.

\begin{figure}
\centerline{\psfig{width=7.8cm,figure=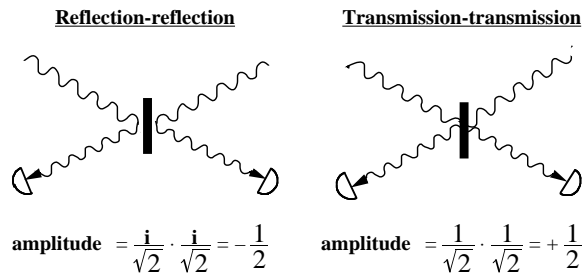}}
\caption{The two indistinguishable processes leading to coincidences}
\label{indist}
\end{figure}
\begin{figure}
\centerline{\psfig{width=10.8cm,figure=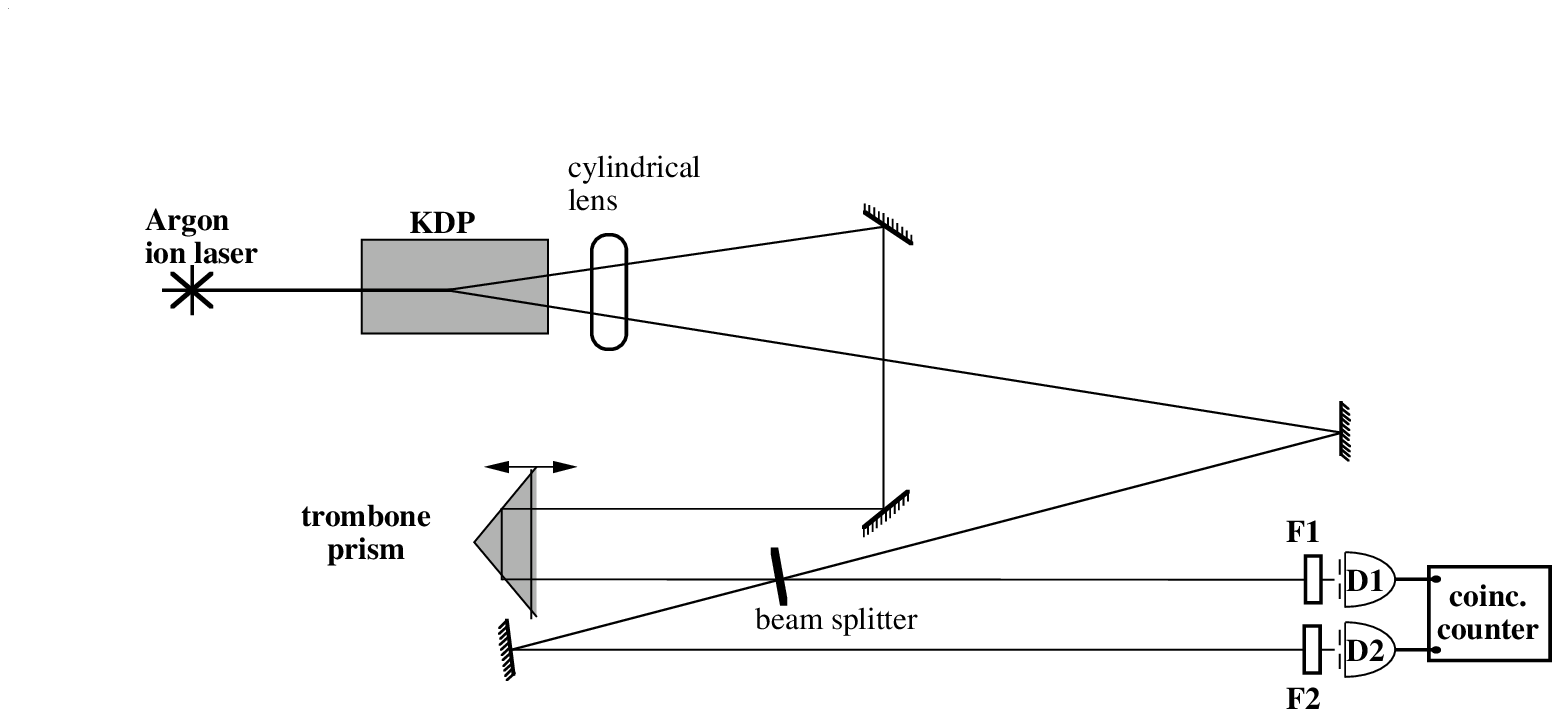}}
\caption{Hong-Ou-Mandel interferometer (Berkeley version)}
\label{ucbhong}
\end{figure}
\begin{figure}
\centerline{\psfig{width=8.2cm,figure=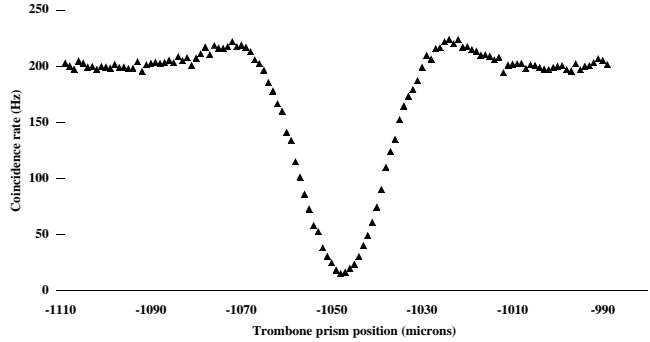}}
\caption{Coincidence rate versus trombone prism position}
\label{trompriz}
\end{figure}
To explain this effect, we again employ Feynman's rules for interference.  The
possible processes for the two photons at the beam splitter are:\\
(1) Both photons are transmitted; the outcome-- a coincidence ``click'' of D1
and D2.\\
(2) and (3) One photon is reflected, the other transmitted;  the outcome-- no
coincidences.\\
(4) Both photons are reflected; the outcome-- a coincidence ``click'' of D1 and
D2.\\
If the two photons reach the beam splitter simultaneously, coincidence
detection processes (1) and (4) are indistinguishable, and thus interfere; see
Fig.~\protect\ref{indist}.  The phase factor $i$ for the reflection amplitude
(relative
to its transmission amplitude) arises from time-reversal symmetry at a
lossless, symmetric beam splitter {\cite{Steinberg=19941D2D}, and results in
destructive interference of the ``reflection-reflection'' and
``transmission-transmission'' probability amplitudes.  Hence the total
amplitude for coincidences to occur is $(1^{2}/2 + i^{2}/2)\!=\!0$:
Coincidences never occur.  In other words, the two photons always exit the same
port of the beam splitter whenever the path length difference is zero, i.e., if
the photon wavepackets arrive at the beam splitter {\it simultaneously}.
However, processes (1) and (4) become distinguishable if the photon wavepackets
arrive at different times at the beam splitter, and coincidences then occur
half the time.  Hence as the path length difference is scanned, one maps out
the overlap of the photon wavepackets; the width of the coincidence dip is a
measure of the coherence length of the single-photon wavepackets.  A schematic
of our version of the HOM interferometer is shown in
Fig.~\protect\ref{ucbhong}.  A
typical coincidence dip is shown in Fig.~\protect\ref{trompriz}.

\begin{figure}
\centerline{\psfig{width=10.8cm,figure=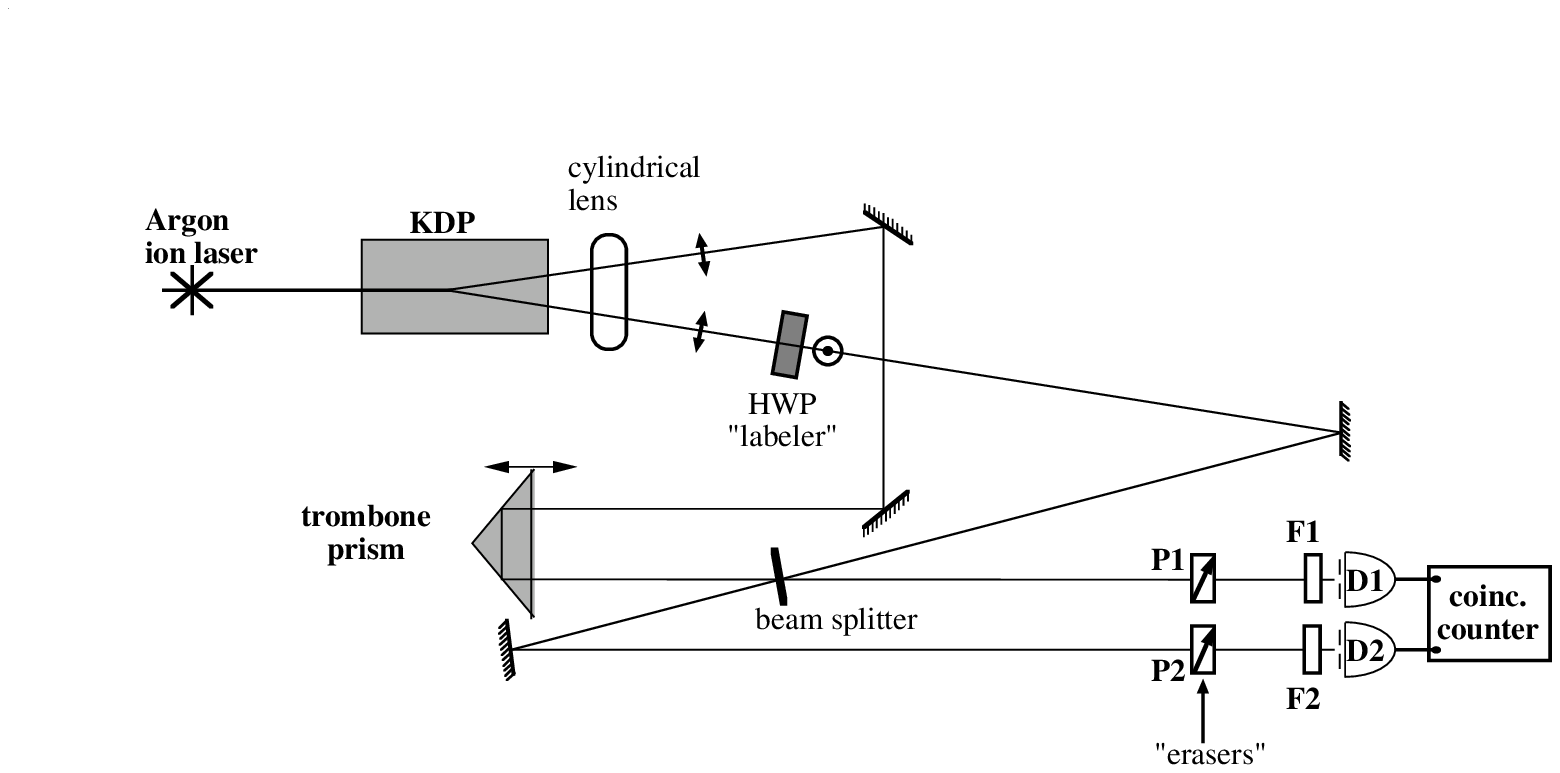}}
\caption{Hong-Ou-Mandel interferometer with ``labeler'' and ``erasers''}
\label{hongle}
\end{figure}
\begin{figure}
\centerline{\psfig{width=8.5cm,figure=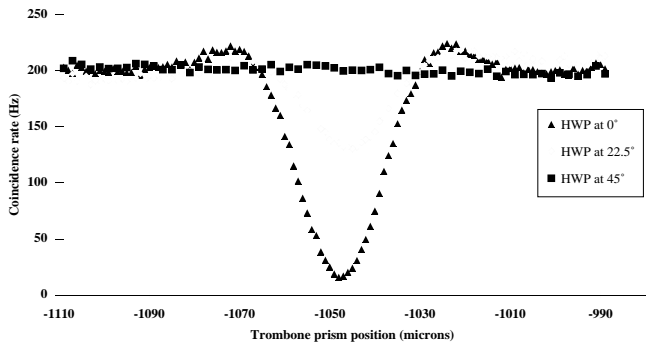}}
\caption{Coincidence rate versus trombone prism position with ``labeler'' in
setup, but without ``erasers''}
\label{coinvstrom}
\end{figure}
\begin{figure}
\centerline{\psfig{width=8.5cm,figure=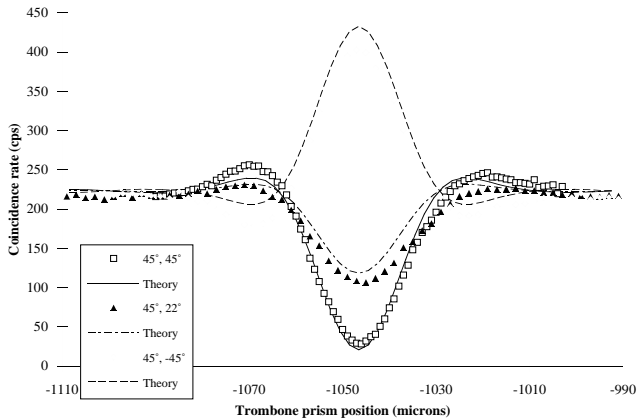}}
\caption{Revival of interference due to (partial) erasure,
for different polarizer settings indicated in the legend.}
\label{revival}
\end{figure}
As in the simpler Young's-type experiment described earlier, in our verson of
the quantum eraser we used polarization as a means of labeling the photons.
The two twin photons emerge from our nonlinear crystal horizontally polarized.
We now add to one arm of the interferometer a half-wave plate (HWP), which can
rotate the photon's polarization to vertical.  This clearly gives information
on which path the photon takes, thereby distinguishing the otherwise
interfering ``reflection-reflection'' and
``transmission-transmission'' processes.  In Fig.~\protect\ref{coinvstrom} we
show the
disappearance of the coincidence dip, as we rotate the the fast axis of the
half-wave plate towards $45^\circ$ with respect to the horizontal, at which
point the polarizations of the down-converted photons are orthogonal, and the
previously interfering paths fully distinguishable. (An intermediate
orientation of the half-wave plate is also shown.)  Now we put in the erasers,
which take the form of two linear polarizers, P1 and P2, in front of the
detectors.  By orienting both of them at $45^\circ$ to the horizontal, we can
erase the which-path information, since both horizontally and vertically
polarized photons end up polarized at $45^\circ$ after passing through these
polarizers, and any possibility of distinguishing between the paths taken by
the photons is lost.  The result is that the interference pattern, i.e., the
coincidence dip, is now revived, as shown by the data represented by the
squares of Fig.~\protect\ref{revival}.  Note that the presence of {\it both}
polarizers
P1 and P2 is necessary to perform the erasure.  The removal of either one would
leave {\it one} of the photons labeled, carrying enough which-path information
to totally destroy the interference pattern.  We stress that it is the mere
{\it possibility} of obtaining which-path information that destroys the
interference; no actual polarization measurements need to be made.  An
interesting feature of this experiment is that one can change the coincidence
dip into a coincidence peak (i.e., an interference minimum into a maximum), by
rotating P1 relative to P2 until one is at $+45^{\circ}$ and the other is at
$-45^{\circ}$.  The data for this orientation are represented by the diamonds
in Fig.~\protect\ref{revival}; an intermediate orientation is also shown.  (We
have
also checked that the coincidence rate at the center of the dip varies
sinusoidally with the {\it relative} angle of P1 and P2, which Shih \& Alley
and Ou \& Mandel have already observed in connection with Bell's inequality
experiments \cite{Shih=1988,Ou=1988PRL}).  In an advanced version of the
quantum eraser experiments, it should even be possible to incorporate a
``delayed choice" feature \cite{Wheeler=1983}, so that the choice of observing
fringes and anti-fringes or retaining which-way information is made {\it after}
the initially interfering particle is detected \cite{Kwiat=19943QE}.

\section{Dispersion Cancellation in Two-Photon Interference}
\label{1.6}

As a motivation for the next experiment, involving dispersion cancellation, let
us return for a moment to the classical problem of propagation in a dispersive
medium.  We know that the peak of a classical electromagnetic wavepacket
propagating through a piece of glass will travel at the group velocity, but it
is not entirely clear that one can interpret this classical wavepacket as if it
were the wavefunction of the single photon and then use the Born interpretation
for this wavefunction.  If this interpretation were correct, then the photon
would simply travel at the group velocity in this medium.  However, as
Sommerfeld and Brillouin have pointed out \cite{Brillouin=1960}, at the
classical level there are at least five kinds of propagation velocities in a
dispersive medium: the phase, group, energy, ``signal,'' and front velocities,
all of which differ from one another in the vicinity of an absorption line,
where there is a region of anomalous dispersion.  In particular, the group
velocity can become ``superluminal,'' i.e., faster than the vacuum speed of
light, in these regions.  If the photon were to travel at the group velocity in
this medium, would it also travel ``superluminally''?  If not, then at which of
these velocities does the photon travel in dispersive media?  (These questions
become especially acute in media with inverted populations, where off-resonance
wavepackets can travel superluminally without attenuation and with little
dispersion \cite{Chiao=1993SUP}; see also the accompanying article by Chiao
{et~al}.)

\begin{figure}
\centerline{\psfig{width=12.85cm,figure=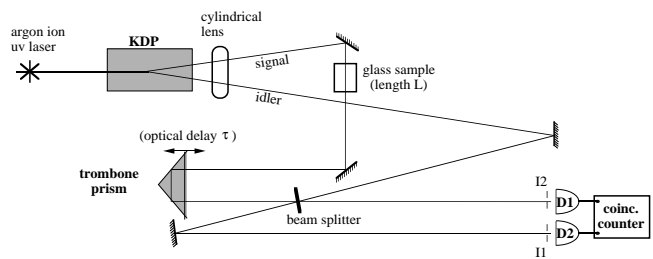}}
\caption{Dispersion cancellation schematic for measuring photon propagation
times: a Hong-Ou-Mandel apparatus with glass sample inserted.}
\label{dispcan}
\end{figure}
Motivated by the above questions, we did the following experiment.  We removed
the HWP and the polarizers from the quantum eraser setup and inserted a piece
of glass in the path of one of the photons; see Fig.~\protect\ref{dispcan}.
The glass
slows down the photon which traverses it, and in order to observe the
coincidence dip, it is necessary to introduce an equal, compensating delay
$\tau$ by adjusting the trombone prism.  We measured the magnitude of
this delay for various samples of glass and were able to determine traversal
times on the order of 35~ps, with $\pm1$~fs accuracy.  In this way, we were
able to confirm that single photons travel through glass at the group velocity
in transparent spectral regions, an interesting example of particle-wave
unity.

\begin{figure}
\centerline{\psfig{width=8.5cm,figure=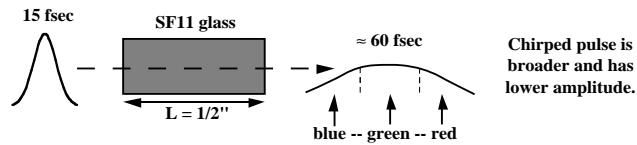}}
\caption{Chirped pulse due to normal dispersion}
\label{chirp}
\end{figure}
Clearly, the interest of measuring optical delays is greatest for media with
dispersion.  Consider the limiting time-resolution of this
interferometer.  For a short wavepacket or pulse, a broad spectrum is
necessary.  In dispersive media, however, the broad spectrum required for an
ultrafast pulse (or single-photon wavepacket) can lead to a great deal of
dispersion.  One might expect that this broadening of the wavepacket would also
broaden the coincidence dip in the HOM interferometer, since the physical
explanation of the dip (in terms of which-path information carried by the
photons' arrival times) seems to imply that the width of the dip should be the
size of the wavepackets which impinge on the beam splitter.  Thus the tradeoff
between pulse width and dispersive broadening would place an ultimate limit on
the resolution of a measurement made on a given sample.  For example, a 15~fs
wavepacket propagating through half an inch of SF11 glass (one of the samples
we studied) would classically broaden to about 60 fs due to the dispersion in
this glass.  The nature of the broadening is that of a chirp, i.e., the local
frequency sweeps from low to high values (for normal dispersion, in which
redder wavelengths travel faster than bluer wavelengths).  Hence the earlier
part of the broadened pulse consists of redder wavelengths, and the later part
of this pulse consists of bluer wavelengths; see Fig.~\protect\ref{chirp}.

In our experiment, however, we found that the combination of the
time-correlations and energy-correlations exhibited by our entangled photons
led to a cancellation of these dispersive effects.  While the individual
wavepacket which travels through the glass does broaden according to classical
optics, it is impossible to know whether this photon was reflected or
transmitted at the beam splitter (recall Fig.~\protect\ref{ucbhong}).  This
means that
when an individual photon arrives at a detector, it is unknowable whether it
travelled through the glass or whether its conjugate (with {\it anticorrelated}
frequency) did so; due to the chirp, the delay in these two cases is opposite,
relative to the peak of the wavepacket.  An exact cancellation occurs for the
(greatly dominant) linear group-velocity dispersion term, and no appreciable
broadening of the 15~fs interference dip occurs.  This is a direct consequence
of the nature of the EPR state, in that it relies on the simultaneous
correlations of energy and time.  A detailed theoretical analysis predicted
these results, in agreement with the simple argument presented here
\cite{Steinberg=1992PRA}.

\begin{figure}
\centerline{\psfig{width=8.5cm,figure=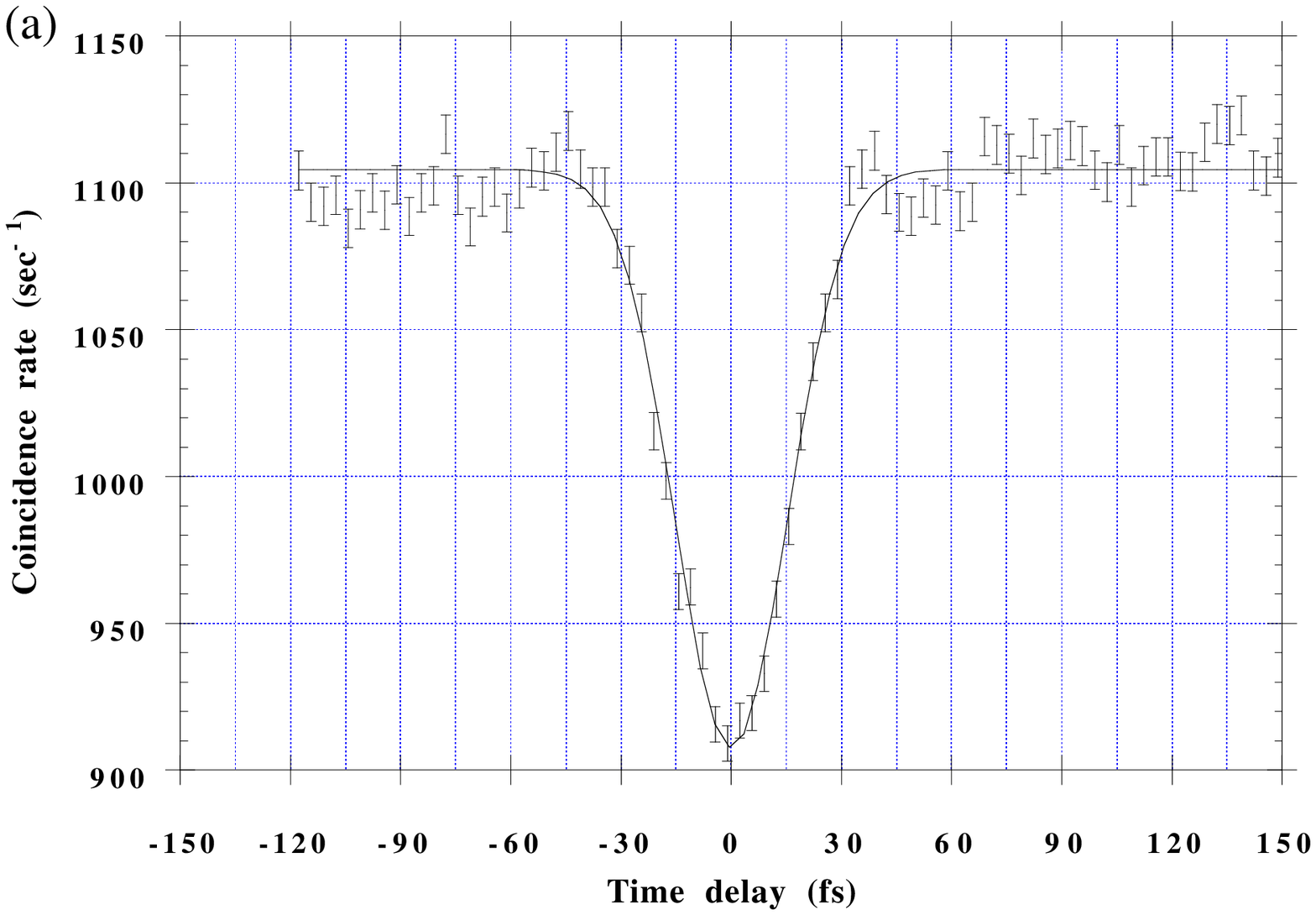}}
\centerline{}
\centerline{\psfig{width=8.5cm,figure=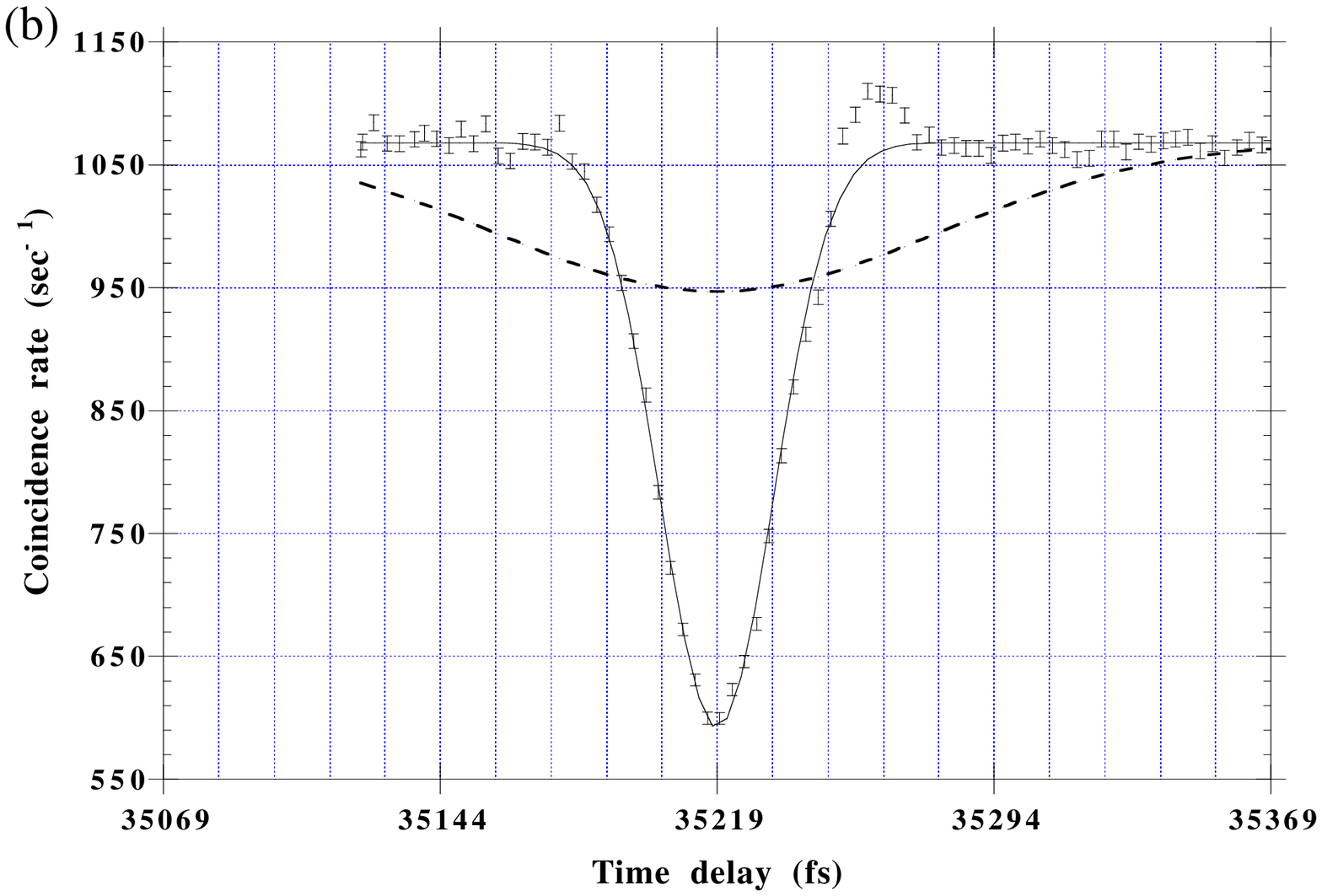}}
\centerline{}
\caption{HOM coincidence rate as a function of the relative optical time delay
in the interferometer. (a) The solid line is a Gaussian fit, with an rms width
of 15.3 fs.  (b) Coincidence profile after a $\frac{1}{2}$-inch piece of SF11
glass is inserted in the signal arm of the interferometer.  The dip shifts by
35,219 fs, but no broadening is observed.  Classically, a 15-fs pulse would
broaden to at least 60 fs, shown for comparison as a dashed line.}
\label{traces}
\end{figure}
In Fig.~\protect\ref{traces}, we see that there is indeed very little
broadening in the
data with the glass compared with that without the glass.  Certainly,
broadening on the scale of 60~fs (the dashed curve) is ruled out by these
data.  Dispersion cancellation is important for applications; particularly, in
the tunneling-time measurements described next, the sharpness of the dip--and
hence the temporal resolution--is not appreciably degraded by the presence of
the dispersion in the optical elements of our apparatus or in the sample
itself.

\section{Measurements of Photon Tunneling Times}
\label{1.7}

Tunneling is one of the most striking consequences of quantum mechanics.  The
Josephson effect in solid state physics, fusion in nuclear physics, and
instantons in high energy physics are all manifestations of this phenomenon.
Every quantum mechanics text treats the calculation of the tunneling
probability.  And yet, the issue of how much {\it time} it takes a particle to
tunnel through a barrier, a problem first addressed in the 1930s, remains
controversial to the present day.  The question arises because the momentum in
the barrier region is imaginary.  The first answer, the group delay (also known
as the ``phase time'' because it describes the time of appearance of a
wavepacket peak by using the stationary phase approximation), can in certain
limits be paradoxically small, implying barrier traversal at a speed greater
than that of light in vacuum \cite{MacColl=1932,Wigner=1955}.  This apparent
violation of Einstein causality does not arise from the use of the
nonrelativistic Schr\"{o}dinger equation, since it also arises in solutions of
Maxwell's equations, which are fully relativistic.  It has generally been
assumed that such superluminal velocities cannot be physical
\cite{Brillouin=1960}, but in the case of tunneling, no resolution has been
universally accepted.

As a result of developments in solid state physics, such as tunneling in
heterostructure devices, the issue has acquired a new sense of urgency since
the 1980s, leading to much conflicting theoretical work
\cite{Buttiker=1982,Hauge=1989,Landauer=1994RMP}.  Several experimental papers
presenting more or less indirect measurements of barrier traversal times have
appeared.  Some seem to agree with the ``semiclassical time'' of B\"{u}ttiker
and Landauer \cite{Buttiker=1982,Landauer=1989}, while others
\cite{Landauer=1993,Stovneng=1993} seem to agree with the group delay (``phase
time'').  We presented the first direct time measurement confirming that the
time delay in tunneling can be superluminal, studying single photons traversing
a dielectric mirror \cite{Steinberg=1993PRL}.  Since then, several microwave
experiments have confirmed that the effective group velocity of classical
evanescent waves in various configurations may be superluminal
\cite{Enders=1993,Nimtz=1994,Ranfagni=1993}.  Also, recently a femtosecond
laser experiment has confirmed our earlier findings of superluminal tunneling
in
dielectric mirrors \cite{Spielmann=1994}, using classical pulses.

Our experiment again employs the down-conversion source in a HOM interferometer
arrangement.  The advantage of using these conjugate particles is that after
one particle traverses a tunnel barrier its time of arrival can be compared
with that of its twin (which encounters no barrier), thus offering a clear
operational definition and direct measurement of the time {\it delay} in
tunneling.  Since this technique relies on coincidence detection, the particle
aspect of tunneling can be clearly observed: Each coincidence detection
corresponds to a single tunneling event.

In our apparatus, the tunnel barrier is a multilayer dielectric mirror.
Such mirrors are composed of quarter-wave layers of alternating high- and
low-index materials, and hence possess a one-dimensional ``photonic band gap''
\cite{Yablonovitch=1991}, i.e., a range of frequencies which correspond to pure
imaginary values of the wavevector.  They are optical realizations of the
Kronig-Penney model of solid state physics, and thus analogous to crystalline
solids possessing band gaps, as well as to superlattices.  Our mirrors
have an $(HL)^{5}H$ structure, where $H$ represents titanium oxide (with an
index of 2.22) and L represents fused silica (with an index of 1.41).  Their
total thickness $d$ is 1.1$\,\mu$m, implying a traversal time of
$d/c\,=\,3.6$~fs if a particle were to travel at $c$.  Their band gaps extend
approximately from 600 to 800~nm, and their transmission amplitudes reach a
minimum of 1\% at 692\,nm.

\begin{figure}
\centerline{\psfig{width=8cm,figure=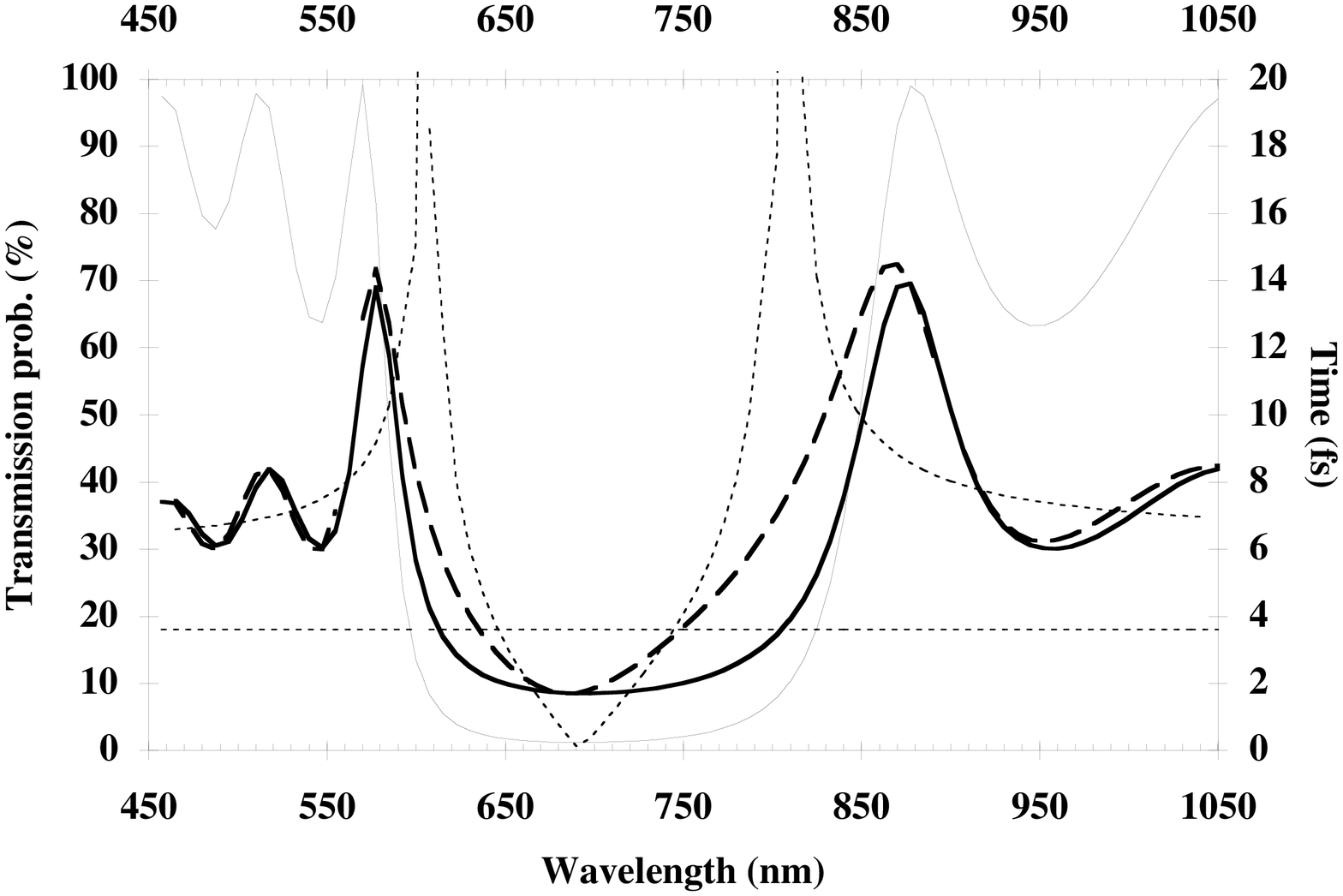}}
\caption{Theoretical curves, where the light solid curve shows the transmission
probability (left axis) of our multilayer coating, as a function of incident
wavelength.  The heavy solid curve shows the group delay, the heavy dashed
curve shows the B\"{u}ttiker-Larmor traversal time, and the light dotted curve
the semiclassical time (right axis).  The horizontal dotted line at 3.6\,fs
represents the ``causality limit'' $d/c$.}
\label{theocurv}
\end{figure}
The semiclassical time is calculated from the group velocity which would hold
inside an infinite periodic medium (i.e., neglecting reflections at the
extremities of the barrier).  As the wavevector becomes pure imaginary for
frequencies within the band gap, so does the semiclassical time; in order to
extend it into the band-gap region, we simply drop the factor of $i$, in
analogy with the interaction time of B\"{u}ttiker and Landauer
\cite{Buttiker=1982}.  The ``Larmor time'' is a measure of the amount of Larmor
precession a tunneling electron would experience in an infinitesimal magnetic
field confined to the barrier region.  B\"uttiker has suggested a Larmor time
which takes into account the tendency of the transmitted electrons to align
their spins {\it along} the magnetic field as well as the precession
{\it about} the field \cite{Buttiker=1983}.  The group delay is the derivative
of the barrier's transmission phase with respect to the angular frequency of
the light, according to the method of stationary phase.  All three times dip
below $d/c\,=\,3.6$~fs and are thus superluminal, although their detailed
behaviors are quite different; see Fig.\,\protect\ref{theocurv}.  For example,
the
group delay remains relatively constant near 1.7~fs over most of the band gap.
The semiclassical time, on the other hand, dips below 3.6~fs only over a
narrower range of frequencies, and actually reaches zero at the center of the
gap.  B\"{u}ttiker's Larmor time approaches the group delay far from the band
gap as well as at its center, but differs from it at intermediate points.

\begin{figure}
\centerline{\psfig{width=8cm,figure=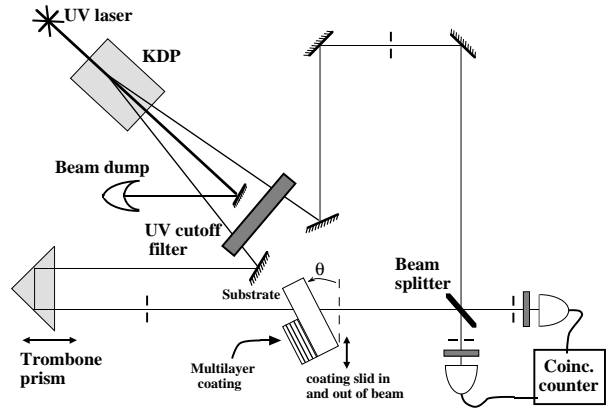}}
\caption{Experimental setup for determining the single-photon propagation times
through a multilayer dielectric mirror.}
\label{angf1}
\end{figure}
Our apparatus is shown in Fig.\,\protect\ref{angf1}.  As before, a KDP crystal
is
pumped by a cw uv laser at 351\,nm, producing pairs of down-conversion photons,
directed by mirrors to impinge simultaneously on the surface of a 50/50 beam
splitter.  One photon of each pair travels through air, while the conjugate
photon impinges on our sample, consisting of an etalon substrate of fused
silica, which is coated over half of one face with the 1.1\,$\mu$m coating
described above, and uncoated on the other half of that face.  (The entire
opposite face is antireflection coated.)  This sample is mounted on two stacked
stages.  The first is a precision translation stage, which can place the sample
in either of two positions transverse to the beam path.  In one of these
positions, the photon must tunnel through the 1.1\,$\mu$m coating in order to
be transmitted, while in the other position, it travels through 1.1\,$\mu$m of
air.  In both positions, it traverses the same thickness of substrate.  The
second stage allows the sample to be tilted with respect to normal incidence.

If the two photons' wavepackets are made to overlap in time at the beam
splitter, the destructive interference effect described above leads to a
theoretical null in the coincidence detection rate.  Thus as the path-length
difference is changed by translating a ``trombone'' prism with a Burleigh
Inchworm system (see Fig.\,\protect\ref{angf1}) the coincidence rate exhibits a
dip
with an rms width of approximately 20\,fs, which is the correlation time of the
two photons (determined by their 6\,nm bandwidths)
\cite{Hong=1987,Steinberg=1992PRA,Jeffers=1993}.  As
explained above, the rate reaches a minimum when the two wavepackets overlap
perfectly at the beam splitter.  For this reason, if an extra delay is inserted
in one arm of this interferometer (i.e., by sliding the 1.1\,$\mu$m coating
into the beam), the prism will need to be translated in order to compensate for
this delay and restore the coincidence minimum.  In order to eliminate so far
as possible any systematic errors, we conducted each of our data runs by slowly
scanning the prism across the dip, while sliding the coating in and out of the
beam periodically, so that at each prism position we had directly comparable
data with and without the barrier.

\begin{figure}
\centerline{\psfig{width=8cm,figure=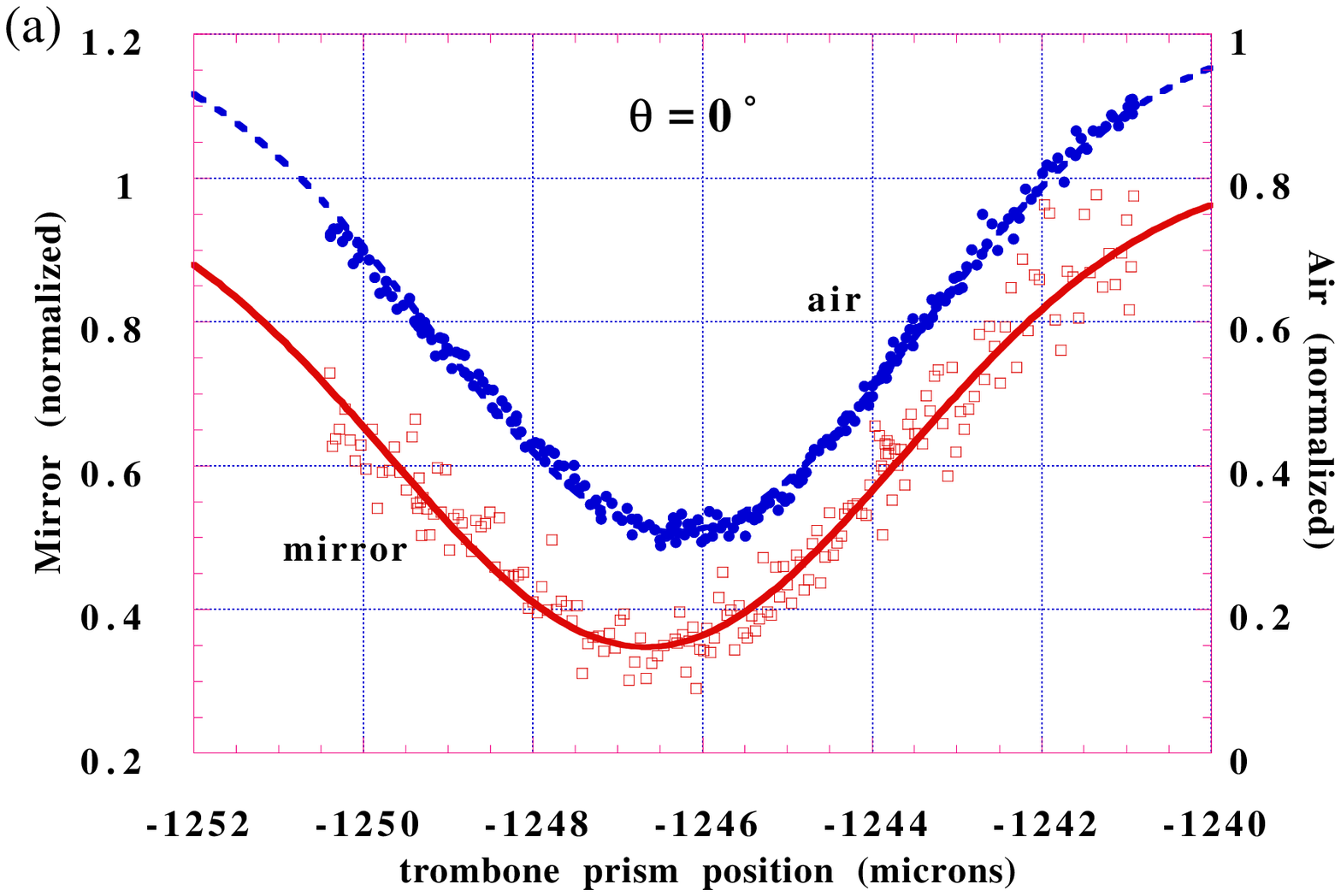}}
\centerline{\psfig{width=8cm,figure=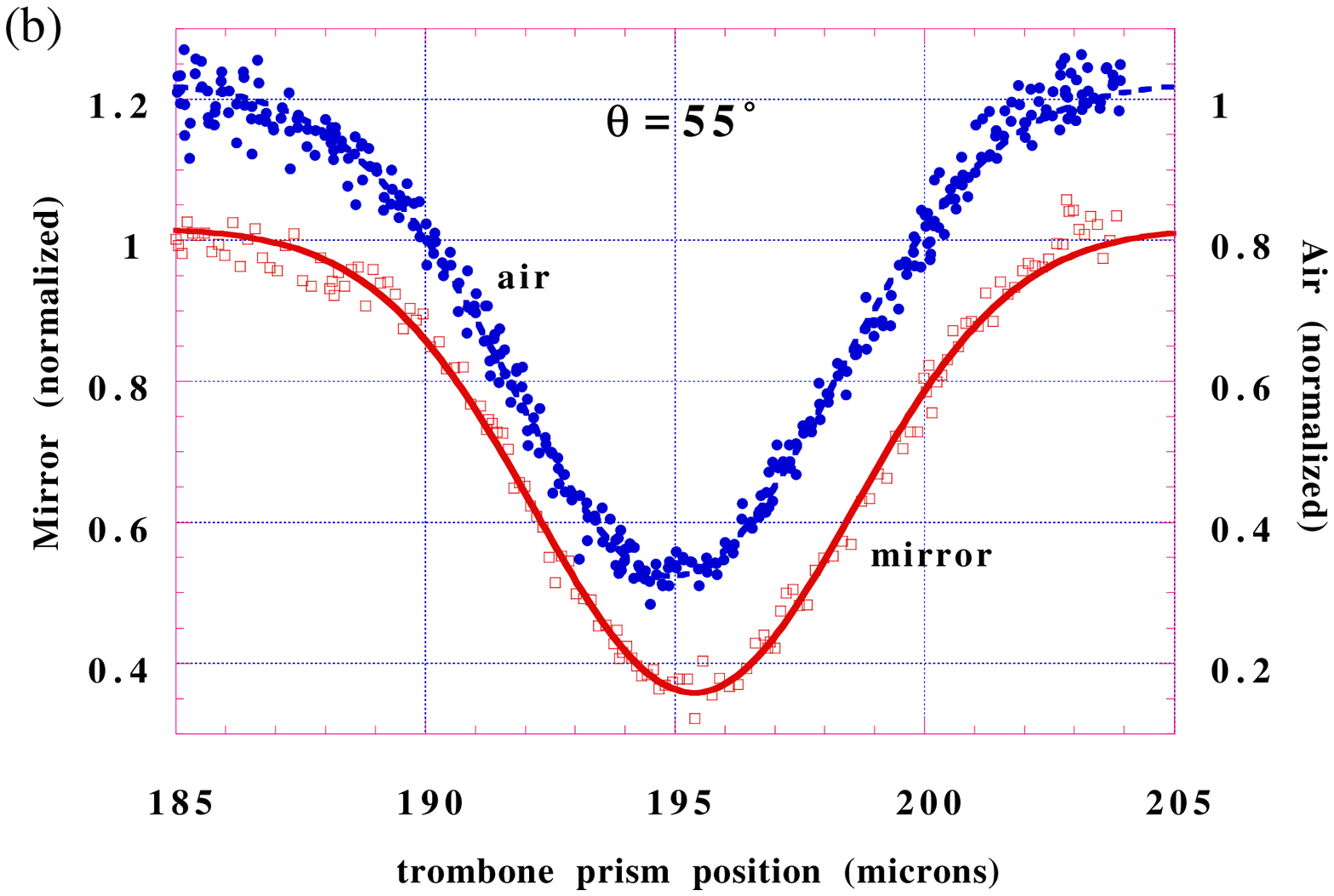}}
\caption{This caption intentionally removed due to latex problems.}
\label{angf2}
\end{figure}
We found that inserting the barrier at normal incidence (for which it was
designed) did in fact cause the dip to be shifted to a position in which the
prism was located {\it farther} from the barrier.  This determines the sign of
the effect: The external delay had to be {\it lengthened}, implying that the
mean delay time experienced by the photon inside the barrier was
{\it less than} the delay time for propagating through the same distance in
air.  As we rotated the mirror about the vertical axis, the bandgap shifted to
lower wavelengths according to Bragg's law, and for the p-polarized photons we
studied, the width of the bandgap also diminished due to the decreased
reflectivity of the dielectric interfaces at non-normal incidence (cf.
Brewster's angle).  Thus at $0^{\circ}$, our 702~nm photons are near the center
of the bandgap, while at $55^{\circ}$, they are near the band edge, where the
transmission is over 40\%.  As can be seen clearly from
Fig.~\protect\ref{angf2}, the
delay time changes from a superluminal value to a subluminal one as the angle
of incidence is scanned, in agreement with theory.

\begin{figure}
\centerline{\psfig{width=8cm,figure=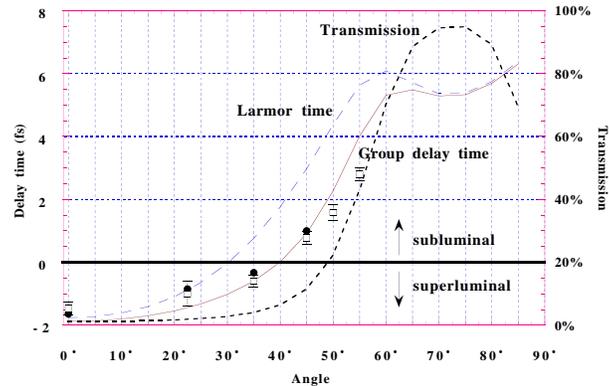}}
\caption{Left axis: measured delay for mirror one ($\Box$) and mirror two
($\bullet$) as a function of angle of incidence, to be compared with the
theoretical group delay and the Larmor interaction time proposed by
B\"{u}ttiker.  Right axis: transmission versus angle of incidence.  All curves
are for p-polarization.}
\label{angf3}
\end{figure}
Our normal-incidence data \cite{Steinberg=1993PRL} demonstrated that the
semiclassical time was inadequate for describing these propagation delays, and
the data were roughly consistent both with the group delay theory and with
B\"{u}ttiker's proposed Larmor time \cite{Buttiker=1983}.  More recent data
\cite{Steinberg=1994SUB} is summarized in Fig.~\protect\ref{angf3}.  As can be
seen,
the evidence for a superluminal delay which becomes subluminal as the photons
approach the band edge is quite convincing.  There is reasonable agreement with
the group delay theory, although a discrepancy on the order of one-half a
femtosecond (and greater at large angles of incidence) remains; we believe this
may be attributable to deviations from an ideal barrier, such as varying layer
thicknesses and residual absorption or scattering.  While the discrepancy is
not yet fully understood, the data for two different mirrors we studied (both
shown in the figure) demonstrate clearly that the observable delay time is
better described by the group delay theory than by the Larmor time.

\begin{figure}
\centerline{\psfig{width=8cm,figure=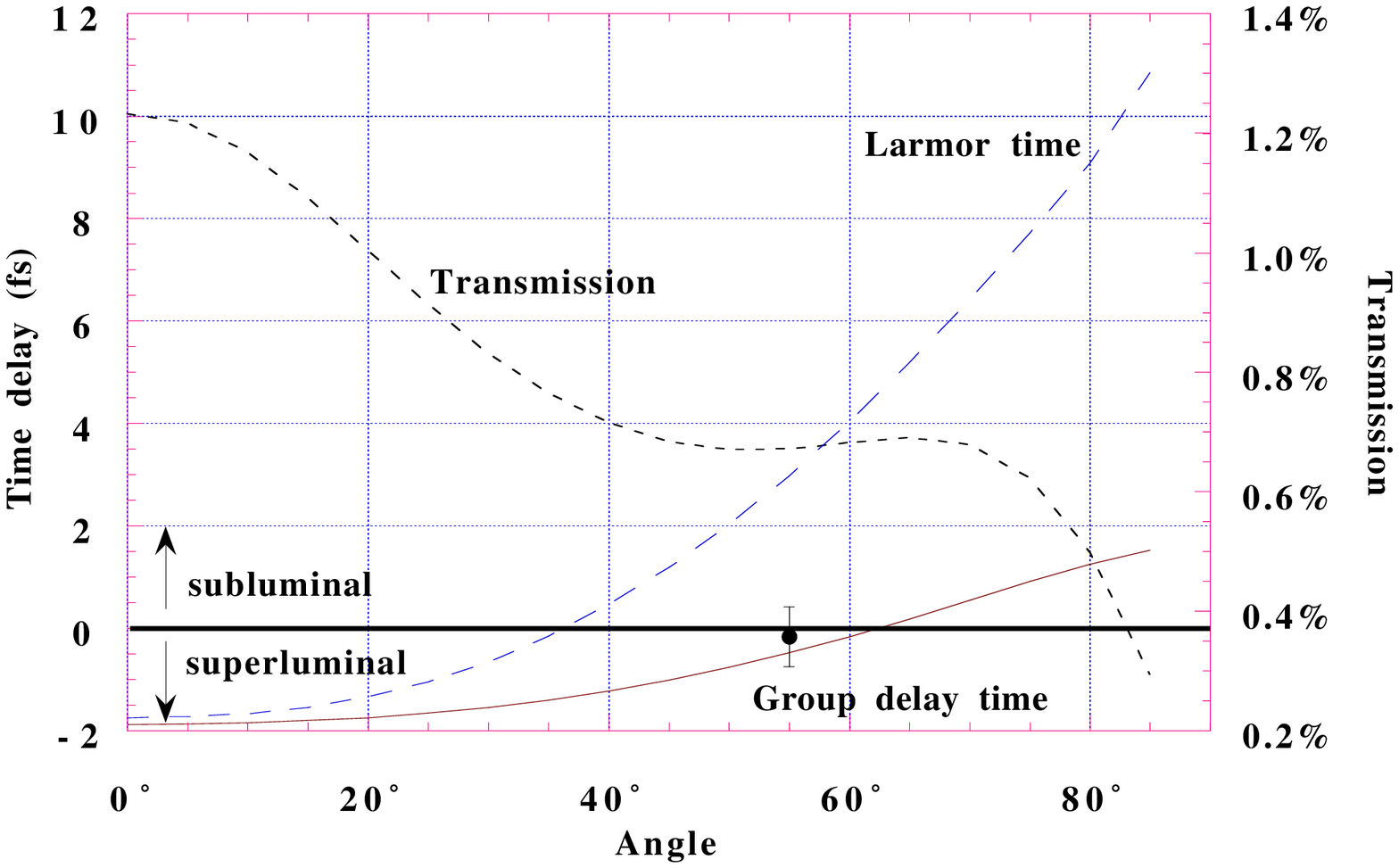}}
\caption{Same as Fig.~\protect\ref{angf3}, but for s-polarization.  Due to the
much
lower transmission, only one preliminary data point is shown, but the different
characteristics of the theoretical curves for both transmission and delay
suggest that upon improvement of our signal-to-noise ratio, further work in
this direction may help elucidate the discrepancy between experiment and
theory (as well as to more conclusively rule out identification of the Larmor
time as descriptive of the measured delay).}
\label{angf4}
\end{figure}
In order to study this discrepancy, in addition to obtaining a second mirror of
the same design parameters, we have begun to measure the delay times for
s-polarization, for which the interface reflectivities increase with angle,
making the bandgap wider instead of narrower.  A preliminary data point is
shown in Fig.~\protect\ref{angf4}, along with theoretical curves which
demonstrate both
why this should be an interesting avenue for further investigation and why it
is more difficult to get an adequate signal-to-noise ratio.  The point shown is
consistent with the group delay, and not with the Larmor time.

We have thus confirmed that the peak of a tunneling wave packet may indeed
appear on the far side of a barrier sooner than if it had been travelling at
the vacuum speed of light.  No signal can be sent with these smooth
wavepackets, however; only a small portion of the leading edge of the incident
Gaussian is actually transmitted, and whether the photon ``collapses'' into
this portion or into the reflected portion is not under experimental control.
Causality is thus not violated by these nonlocal effects.  The superluminality
can be understood by thinking of the low transmission through our barrier as
arising from destructive interference between waves which have spent different
lengths of time in the barrier.  While the incident wavepacket is rising,
multiple reflections can be neglected, since their intensities are small
relative to the partial wave which makes a single pass; thus the destructive
interference is not very effective.  At later times, when the fields stored in
the barrier have had time to reach a steady state, the interference reduces the
transmission to its steady-state value.  Thus the leading edge of the packet is
transmitted preferentially with respect to the rest of the packet, shifting the
transmitted peak earlier in time.

Recent work based on ``weak measurement'' theory \cite{Aharonov=1988} and the
idea of conditional probability distributions for the position of a quantum
particle suggests that this superluminal effect is related to the fact that a
tunneling particle spends very little time in the barrier region, except
within an evanescent decay length of the two barrier edges
\cite{Steinberg=1994WK,Steinberg=1994PRA,Steinberg=1994TH}.  It is as though
the particle ``skipped'' the bulk of the barrier.  Furthermore, the nonlocality
is underscored by the fact that this approach allows one to describe
conditional probability distributions for a particle which is first prepared
incident on the left and later detected emerging on the right.  These
probability distributions describe in-principle measurable effects, and do
indeed traverse the barrier faster than the vacuum speed of light.
They suggest that a single tunneling particle could affect the expectation
values of two different measuring devices located at spacelike separated
positions, so long as the coupling to the devices was too weak to disturb
the tunneling process, and hence too weak to shift either measuring device by
an amount comparable to its intrinsic uncertainty.

\section{Conclusion}

The experiments which we have described in this paper demonstrate some of the
stranger nonlocal features of quantum mechanics.  The first three of these
experiments explore them in connection with the Einstein-Podolsky-Rosen
effect.  In the Franson experiment, the behaviors of the two space-like
separated particles at the final beam splitters (i.e., which exit port they
choose) are correlated or anticorrelated with each other, depending on the
settings of the phase shifters in the interferometer.  Likewise, in the
quantum eraser, whether interference or the complementary which-path
information is observed can be controlled by the experimenter's choice of the
settings of polarizers placed after the final beam splitter of the
interferometer.  In the dispersion cancellation experiment, one cannot know,
even in principle, which of two photons propagated through a piece of the
glass.  This in turn leads to a cancellation of the effect of dispersive
broadening on the measurement.  The fourth of these experiments shows that,
even at the one-particle level, there exist nonlocal effects in quantum
mechanics: in tunneling there exist superluminal time delays of the tunneling
particle.  None of these nonlocal effects violates Einstein causality, due to
the uncontrollable randomness of quantum events.  In the fourth experiment,
there is another way to understand that Einstein causality is not violated;
the front velocity, at which {\it discontinuities} propagate, never exceeds
$c$ (see accompanying article by Chiao {\it et~al}).

\section{Acknowledgments}

We would like to thank Jack Boyce, Jeff Chamberlain, Josh Holden, Bruce
Johnson, Grant McKinney and Morgan Mitchell for their assistance in the
tunneling time experiment and preparation of this manuscript, and
Winfried Heitmann for helpful e-mail.  RYC would like to thank
Prof. S. Zubairy for an invitation to speak at the '94 summer school in
Nathiagali, Pakistan at which these results were presented.  This work was
supported by ONR grant number
\mbox{N00014-90-J-1259}.


\begin{thebibliography}{1}

\bibitem{Einstein=1935}
Einstein A, Podolsky B and Rosen N 1935 {\it Phys. Rev.} {\bf 47} 777

\bibitem{Franson=1989}
Franson J D 1989  {\it Phys. Rev. Lett.} {\bf 62} 2205

\bibitem{Kwiat=1993FR}
Kwiat P G, Steinberg A M and Chiao R Y 1993 {\it Phys. Rev.} A {\bf 47} R2472

\bibitem{Kwiat=1992}
Kwiat P G, Steinberg A M and Chiao R Y 1992 {\it Phys. Rev.} A {\bf 45} 7729

\bibitem{Steinberg=1992PRL}
Steinberg A M, Kwiat P G and Chiao R Y 1992 {\it Phys. Rev. Lett.} {\bf 68}
2421

\bibitem{Steinberg=1993PRL}
Steinberg A M, Kwiat P G and Chiao R Y 1993 {\it Phys. Rev. Lett.} {\bf 71} 708

\bibitem{Steinberg=1994SUB}
Steinberg A M and Chiao R Y 1994 {\it Phys. Rev.} Submitted (quant-ph/9501013)
{\it Sub-femtosecond determination of transmission delay times for a dielectric
mirror (photonic bandgap) as a function of angle of incidence}

\bibitem{Bohm=1983}
Bohm D 1983 in {\it Quantum Theory and Measurement}
ed J A Wheeler and W H Zurek (Princeton: Princeton) p 356

\bibitem{Freedman=1972}
Freedman S J and Clauser J F 1972 {\it Phys. Rev. Lett.} {\bf 28} 938

\bibitem{Clauser=1978}
Clauser J F and Shimony A 1978 {\it Rep. Prog. Phys.} {\bf 41} 1881

\bibitem{Aspect=1982ADR}
Aspect A, Dalibard J and Roger G 1982 {\it Phys. Rev. Lett.} {\bf 49} 1804

\bibitem{Bell=1964}
Bell J S 1964 {\it Physics} {\bf 1} 195

\bibitem{Rarity=1990BI}
Rarity J G and Tapster P R 1990 {\it Phys. Rev. Lett.} {\bf 64} 2495

\bibitem{Schrodinger=1983}
Schr\"{o}dinger E 1983 in {\it Quantum Theory and Measurement}
ed J A Wheeler and W H Zurek (Princeton: Princeton) p 152

\bibitem{Chiao=1991MD}
Chiao R Y, Kwiat P G and Steinberg A M 1991 {\it Proceedings Workshop on
Squeezed States and Uncertainty Relations} ed D Han, Y S Kim and W W Zachary
(NASA Conference Publication 3135) p 61

\bibitem{Brendel=1992}
Brendel J, Mohler E and Martienssen W 1992  {\it Europhys. Lett.} {\bf 20} 575

\bibitem{Rarity=1994}
Rarity J G and Tapster P R 1994 {\it Phys. Rev. Lett.} {\bf 73} 1923

\bibitem{Clauser=1969}
Clauser J F, Horne M A, Shimony A and Holt R A 1969 {\it Phys Rev. Lett.}
{\bf 23} 880

\bibitem{Kwiat=1993TH}
Kwiat P G 1993 {\it Nonclassical Effects from Spontaneous Parametric
Down-Conversion: Adventures in Quantun Wonderland}
PhD thesis (U~C~Berkeley)

\bibitem{Scully=1991}
Scully M O, Englert B -G and Walther H 1991 {\it Nature} {\bf 351} 111

\bibitem{Feynman=1965}
Feynman R P, Leighton R B and Sands M 1965 {\it The Feynman Lectures on
Physics} (Reading, MA: Addison-Wesley) {\bf III} 3-5

\bibitem{Stern=1990}
Stern A Aharonov Y and Imry Y 1990 {\it Phys. Rev.} A {\bf 41} 3436

\bibitem{Jordan=1993}
Jordan T F 1993 {\it Phys. Rev.} A {\bf 48} 2449

\bibitem{Hong=1987}
Hong C K, Ou Z Y and Mandel L 1987 {\it Phys. Rev. Lett.} {\bf 59} 2044

\bibitem{Steinberg=19941D2D}
Steinberg A M and Chiao R Y 1994 {\it Phys. Rev.} A {\bf 49} 3283

\bibitem{Shih=1988}
Shih Y H and Alley C O 1988 {\it Phys. Rev. Lett.} {\bf 61} 2921

\bibitem{Ou=1988PRL}
Ou Z Y and Mandel L 1988 {\it Phys. Rev. Lett.} {\bf 61} 50

\bibitem{Wheeler=1983}
Wheeler J A 1983 in {\it Quantum Theory and Measurement}
ed J A Wheeler and W H Zurek (Princeton: Princeton) p 182

\bibitem{Kwiat=19943QE}
Kwiat P G, Steinberg A M and Chiao R Y 1994 {\it Phys. Rev. A}
{\bf 49} 61

\bibitem{Brillouin=1960}
Brillouin L 1960 {\it Wave Propagation and Group Velocity} (New York: Academic
Press)

\bibitem{Chiao=1993SUP}
Chiao R Y 1993 {\it Phys. Rev. A} {\bf 48} R34

\bibitem{Steinberg=1992PRA}
Steinberg A M, Kwiat P G and Chiao R Y 1992 {\it Phys. Rev. A} {\bf 45} 6659

\bibitem{MacColl=1932}
MacColl L A 1932 {\it Phys. Rev.} {\bf 40} 621

\bibitem{Wigner=1955}
Wigner E P 1955  {\it Phys. Rev.} {\bf 98} 145

\bibitem{Buttiker=1982}
B\"{u}ttiker M and Landauer R 1982 {\it Phys. Rev. Lett.} {\bf 49} 1739

\bibitem{Hauge=1989}
Hauge E H and St{\o}vneng J A 1989 {\it Rev. Mod. Phys.} {\bf 61} 917

\bibitem{Landauer=1994RMP}
Landauer R and Martin T 1994 {\it Rev. of Mod. Phys.} {\bf 66} 217

\bibitem{Landauer=1989}
Landauer R 1989 {\it Nature} {\bf 341} 567

\bibitem{Landauer=1993}
Landauer R 1993  {\it Nature} {\bf 365} 692

\bibitem{Stovneng=1993}
St{\o}vneng J A and Hauge E H 1993  {\it Phys. World} {\bf 6} 23

\bibitem{Enders=1993}
Enders A and Nimtz G 1993 {\it J. Phys. I} {\bf 3} 1089

\bibitem{Nimtz=1994}
Nimtz G, Enders A and Spieker 1994 {\it J. Phys. I} {\bf 4} 565;
Steinberg A M 1994 {\it J. Phys. I} {\bf 4} 1813

\bibitem{Ranfagni=1993}
Ranfagni A,Fabeni P, Pazzi G P, and Mugnai D 1993 {\it Phys. Rev. E} {\bf 48}
1453

\bibitem{Spielmann=1994}
Spielmann Ch, Szip\"{o}cs R, Stigl A and Krausz F 1994 {\it Phys. Rev. Lett.}
{\bf 73} 2308

\bibitem{Yablonovitch=1991}
Yablonovitch E and Leung K M 1991 {\it Physica} {\bf 175B} 81 and references
therein

\bibitem{Buttiker=1983}
B\"{u}ttiker M 1983 {\it Phys. Rev.} B {\bf 27} 6178

\bibitem{Jeffers=1993}
Jeffers J and Barnett S M 1993 {\it Phys. Rev.} A {\bf 47} 3291

\bibitem{Aharonov=1988}
Aharonov Y and Vaidman L 1988 {\it Phys. Rev. Lett.} {\bf 58} 1351

\bibitem{Steinberg=1994WK}
Steinberg A M 1994 {Phys. Rev. Lett.} submitted (quant-ph/9501015)
{\it How much time does a tunneling particle spend in the barrier region?}

\bibitem{Steinberg=1994PRA}
Steinberg A M 1994 {\it Phys. Rev.} A submitted
{\it Conditional probabilities in quantum theory, and the tunneling time
controversy}

\bibitem{Steinberg=1994TH}
Steinberg A M 1994 {\it When Can Light Go Faster Than Light? The
tunneling time and its sub-femtosecond measurement via quantum
interference} PhD thesis (U~C~Berkeley)

\end{thebibliography}
\end{document}